\documentclass[12pt,preprint]{aastex}
\def\lsim{\lower.5ex\hbox{$\; \buildrel < \over \sim \;$}}
\def\gsim{\lower.5ex\hbox{$\; \buildrel > \over \sim \;$}}

\begin{document}

\title{Spectrum of two component flows around a super-massive black hole: an application to M87} 

\author {Samir Mandal\altaffilmark{1} and Sandip K. Chakrabarti\altaffilmark{2,1}}

\altaffiltext{1}{Indian Centre for Space Physics, Chalantika 43, Garia Station Rd.,\\
Garia, Kolkata, 700084, e-mail: samir@csp.res.in}
\altaffiltext{2}{ S.N. Bose National Centre for Basic Sciences,\\
JD Block, Salt Lake, Sector III, Kolkata 700098, e-mail: chakraba@bose.res.in}

\begin{abstract}
We calculate the spectra of two-component accretion flows around black holes of 
various masses, from quasars to nano-quasars.
Specifically, we fit the observational data of
M87 very satisfactorily using our model and find that the 
spectrum may be well fitted by a sub-Keplerian component alone, and
there is little need of any Keplerian component. The non-thermal distribution of 
electrons produced by their acceleration across the standing shock in the sub-Keplerian 
component is enough to produce the observed flat spectrum through the synchrotron radiation.
\end{abstract}


\section{Introduction}
The general view of the physics of the quasars and active galactic nuclei (AGN) is that their energy output is due to 
accretion of matter on to a massive black hole (e.g., Blandford 1991; Antonucci 1993 and references therein). 
Analysis of the spectrum is very important in understanding the physical nature of the underlying flow. 
The observed spectral and timing properties of galactic black holes indicate that the accretion 
flow around a black hole may have two components (Smith et al. 2001; Smith et al. 2002; Choudhury \& Rao 2004; 
Pottschmidt et al. 2006; Smith et al. 2007): 
an optically thick and geometrically thin Keplerian accretion disk (Shakura \& Sunyaev 1973) on 
the equatorial plane and an optically thin sub-Keplerian flow (Chakrabarti \& Titarchuk 1995; 
Chakrabarti \& Mandal 2006) sandwiching the Keplerian disk. The Keplerian disk produces a 
multi-colour black body spectrum and the soft photons from the Keplerian disk are inverse Comptonized 
by a hot electron cloud close to the black hole. In the literature many proposals have been put forward
for the Comptonizing region (Sunyaev \& Titarchuk 1980; Sunyaev \& Titarchuk 1985; Haardt 1994; 
Poutanen \& Svensson 1996) which are
ranging from hot Compton cloud to corona above accretion Keplerian disks.
In a departure from this approach, Chakrabarti \& Titarchuk (1995) proposed that both the so-called
Compton cloud and soft-photon supplier are in fact dynamically important components. 
This model has a hot region produced due to a shock transition in presence of the centrifugal force 
in the sub-Keplerian flow and is known as the CENBOL (CENtrifugal pressure 
supported BOundary Layer). In the present model, the CENBOL is responsible for producing the high-energy 
spectrum from an accretion disk. This two-component nature of the accretion disk can be treated as a general 
model where the accretion rates of both the components can be varied 
independently and the model should be applicable for all the black hole candidates
(belonging to the usual quasars and AGNs to nano-quasars or stellar mass black holes).
It is well known that the big-blue bump in an AGN spectrum is 
the signature of the Keplerian disk and a number of spectra from the core of AGNs have been 
successfully fitted (Sun \& Malkan 1989) by a standard disk model. 
On the other hand, an object such as M87 does not seem to have the big-blue bump in its spectrum and one 
should require only a hot electron component, such as a sub-Keplerian flow, to interpret the observed spectrum.
Fortunately, the same CENBOL component is capable of launching an outflow (Chakrabarti, 1999) which manifests
itself as a strong jet coming out of a region from within a few tens of Schwarzschild radii 
(Junor, Biretta \& Livio, 1999). The predicted profiles of the lines emitted 
from the sub-Keplerian component also seem to agree with observed profiles from HST (Chakrabarti, 1995).

In this {\it Letter}, we calculate the spectra of two-component accretion disks around
stellar mass to super-massive black holes. This helps us to 
understand the basic functions of the components of the flow from an unifying view by varying only the mass. 
We then apply our model to the case of M87 by considering a purely sub-Keplerian disk around it.
The paper is organized in the following way: In \S 2, we describe the model properties
and the general nature of the spectrum. In \S 3, we explain the spectrum of M87 and finally, 
in \S 4, we present our concluding remarks.

\section{Model properties and the general nature of the spectrum from a two-component flow}

We consider a vertically averaged two-component flow around a Schwarzschild black hole.
The black hole geometry is described by a pseudo-Newtonian potential (Paczy\'nski \& Wiita 1980) and the 
vertical height of the accretion disk at any radial distance has been calculated by balancing the vertical component of 
gravitational force with the gas pressure (Chakrabarti 1989). The radial distance is measured in units of the
Schwarzschild radius ($r_g=2GM/c^2$), where $G$ is gravitational constant, $M$ is mass of the black hole 
and $c$ is speed of light. The parameters of our model are: the shock location ($x_s$), the compression ratio ($R$) 
of the shock (i.e., the ratio of the post-shock to pre-shock densities),
the fraction of electrons having non-thermal distribution of energy ($\xi$) and the accretion rates 
of the Keplerian (${\dot m}_d$) and the sub-Keplerian halo (${\dot m}_h$) components. 
The accretion rates are measured in units of the Eddington rates ($0.2 M_\odot/year$ for $10^8 M_\odot$ black hole). 
In the absence of a satisfactory description of magnetic fields inside the accretion disk, 
we consider the presence of only stochastic fields and use the
ratio of the magnetic field energy density to the gravitational energy density to be a parameter ($\beta$). We have taken 
$\beta=1.0$ throughout this paper.
The condition of the plasma at the outer boundary of the accretion disk is uncertain
because the processes of feeding a black hole are not well-known. It depends on the type of the donor star 
(Bath \& Shaviv 1976) supplying matter in a compact binary or the condition of the ISM for AGN. 
We have fixed the outer boundary at 
$x_{in}=10^6 r_g$. The injection temperatures of the electron and proton for stellar-mass black hole
are given by the adiabatic Bondi (1952) solution i.e., $T_{in}=1.6\times10^6 K$ which is consistent with the value found in the 
literature (Tavani \& Brookshaw 1995). However, for AGNs, the outer 
boundary of the accretion disk could be cooler than that around a stellar-mass 
black hole as it contains the molecular or partially ionized gas. The temperature 
of the ionized gas lies in the range $(1.0 - 1.7)\times10^4 K$ (Wilson \& Storchi-Bergmann 1997). 
As a example, we have taken the temperature at the outer boundary for a super-massive black hole as $T_{in}=1.6\times10^4 K$. 
In the case of M87, the existence of an ionized disk has been discussed in the literature (Ford et al. 1994; Harms et al. 1994).
In our model, we consider all the radiation processes, namely bremsstrahlung, synchrotron and Compton scattering.
Since the shocks are the natural outcome of a sub-Keplerian flow,
we consider that the pre-shock electrons follow a pure thermal distribution while the post-shock flow is a
mixture of thermal and non-thermal electrons. This is because some electrons would be accelerated due to
usual back and forth diffusion and compression across the shock (see, Mandal \& Chakrabarti 2005 and references therein). 
The slope ($p=\frac{R+2}{R-1})$ of the non-thermal distribution depends on the compression ratio 
($R$) and it produces a synchrotron spectrum of power-law 
index [$\alpha=(1-p)/2$] with a sharp cut-off determined by the Lorentz factor ($\gamma$) of the accelerated electrons. 
We have taken this effect also in calculating the spectrum from the accretion disk. 

AGNs are known to have very strong jets and in the core/nuclear region these jets are not separable from 
the accretion disk. So, the spectrum from the core will always have some contribution from the jet. 
There could be several hot regions which emit radio, optical and x-ray radiation along the jet but these regions are
separable from the nucleus. Understanding these high energy emissions require separate physical processes
which are outside the scope of the present work.
In our present calculation we are mainly interested the spectrum from the nucleus and have used a simplified 
model for the jet and added this contribution to the spectrum from the accretion disk. 
We have assumed that 10\% of matter from the accretion disk is launched from the 
location of the CENBOL as jet with the same temperature as that of CENBOL. We have taken a cylindrical jet and solve the
energy equations along the jet. The jet is assumed to cool down due to synchrotron emission.  

In Fig. 1 we show a typical accretion disk spectrum for a  stellar 
mass black hole (the so-called nano-quasar) to a super-massive black hole (quasar) with all the
components computed from our model. The parameters are: $x_s=20.0, R=2.5, \xi=0.01, \gamma=2.7\times10^2, 
{\dot m}_d=0.5, {\dot m}_h=0.1$.
The contribution from the bremsstrahlung radiation is negligible for a stellar-mass black hole but it is significant 
for a super-massive black hole due to a lower temperature and larger emitting volume of the accretion disk. The  
pre-shock synchrotron radiation is huge for a stellar-mass black hole due to the large injection temperature whereas 
it is insignificant for a super-massive black hole. The Comptonized spectrum of the black body 
photons gets harder as the black hole mass increases.
This is due to fact that for the same value of accretion rate (in Eddington unit) the density of the flow goes down
with the mass and hence the optical depth of the flow decreases. But the Comptonized spectrum of the synchrotron photons
becomes softer with the increase of the black hole mass. The ratio ($\epsilon$) of the photon energy density to the
magnetic energy density represents the relative contribution of the inverse-Compton process. 
We find that for the stellar mass black hole $\epsilon=0.26$ whereas for the  super-massive case $\epsilon=0.01$ only. 
The jet, which is emitted from CENBOL, 
can have a big contribution in the radio range for the super-massive case due to its large volume. 
So, the radio emission from the nucleus of the AGN will be contributed by the jets. 
We also note a frequency shift towards the lower end as the potential energy release itself decreases with increasing 
mass. Though generally speaking, the spectrum of a super-massive black hole will have all the components of that of 
a stellar mass black hole, we show below that M87 could be fitted by considering a pure sub-Keplerian flow as
it does not seem to have a so-called Big Blue Bump. This was also noted by Perlman et al. (2001). 

\section{Application to M87}

The elliptical galaxy M87 contains a super-massive black hole of mass $M = (3.2 \pm 0.9)\times 10^9 M_\odot $ 
(Macchetto et al. 1997) at the centre and the inclination angle of the accretion disk with the line of sight is 
$ i= (42 \pm 5)^\circ$ (Ford et al. 1994; Chakrabarti, 1995). It is a low luminosity AGN located in the Virgo cluster
at a distance of $ D = (16 \pm 1.2) Mpc $ (Tonry et al. 2001) having a prominent one-sided jet. 
The central luminosity of the accretion disk is $\sim 10^{42}$ ergs/s (Biretta, Stern \& Harris, 1991) 
which is at least two orders of magnitude below the luminosity expected for a standard thin 
accretion disk accreting at the Bondi rate $\dot M_B = 0.1 M_\odot \ yr^{-1}$ 
(Di Matteo et al. 2003). In the literature, different explanations for the low luminosity of M87 have been given.
The wave-particle resonance can efficiently couple the electrons and
ions (Begelman \& Chiueh 1988; Bisnovatyi-Kogan \& Lovelace 2000; Quataert 1998; Quataert \& 
Gruzinov 1999; Blackman 1999) to produce a geometrically thin cool disk.  
Jolley \& Kuncic (2007) argued that this low luminosity is due to a thin cool disk accreting matter in a 
very low rate. On the other hand, if the coupling between electrons and ions are unable to equilibrate 
them within the infall time-scale and if the ions are preferentially heated by viscous dissipation, 
the accreting matter cannot radiate its internal energy before reaching the black hole. This leads to 
a Radiatively Inefficient Accretion Flow (RIAF) (Narayan \& Yi 1994). According to this model the low 
luminosity is due to low radiative efficiency (Di Matteo et al. 2000) rather than a low mass accretion rate. 
But our approach is different from the above two in the sense that our accretion disk is neither a cool Keplerian 
disk because the observation data shows a flat spectrum rather than a blue bump nor a RIAF. Ours is 
simply a sub-Keplerian transonic flow (Chakrabarti, 1990) which is equipped with a standing shock wave.
This is fundamentally inefficient as the infall time is too short. 
The observed luminosity from the nucleus is likely to be less
than the total accretion power because most of the energy is carried inside the 
black hole (In fact, a perfectly stable solution will exist even if the energy loss is zero.)
and rest is used to power the jet. In the case of M87 the total kinetic power of the jet
is estimated to be $\sim 2 \times 10^{43}$ ergs/s (Reynolds et al. 1996). 
Also, the jet is produced from a central region not more than a few tens of  $r_g$ (Junor, Biretta \& Livio 1999) 
which is expected from our CENBOL paradigm for the origin of jet (Chakrabarti, 1999).
There are several works in the literature which favor the sub-Keplerian flow scenario.
For example, it is believed that the low-ionization nuclear emission-line regions 
(LINER) of M87 are produced due to the shock excitation in a dissipative accretion (Dopita et al. 1997). 
Chakrabarti (1995) calculated the velocity field of the ionized disk using a spiral shock solution and comparing the 
shapes of the line profiles expected from various regions of the ionized disk with the HST observation data, the 
mass of the central object of M87 was found to be $M = (4.0 \pm 0.2)\times 10^9 M_\odot$ which is consistent 
with the presently accepted value.

We have collected the broadband (radio, optical to x-ray) data from several previous works (Ho, 1999; Reimer et al. 2004) 
in the literature. The details are given in Table 1. In Fig. 2, we fitted these data by our model.  
We chose the parameters to be $M = 3.2\times 10^9 M_\odot,  x_s= 10.0, R=2.5, \gamma=2.7\times10^2,   
\xi=0.006, {\dot m}_h=0.3$ to fit the data. The upper limit of the Keplerian disk rate is found to be 
${\dot m}_d = 0.001$ for any decent fit. This indicates that the Keplerian disk is not important for M87 
and the bulk motion Comptonization effect is negligible due to the small accretion rate required.
The jet has a large contribution in the radio range due to high electron temperature and large volume. 
The low energy data are well fitted by the thermal synchrotron radiation from the jet and the synchrotron emission   
produced by the cool pre-shock flow is insignificant. But the radio data (open triangles) 
which have a very high spatial resolution are very close to the pre-shock synchrotron spectrum. This radio contribution
may be due to the accretion disk or it may be due the different observation epoch when the radio activity in the jet 
is very dim. The bremsstrahlung radiation has a small contribution only in the soft x-ray range. 
A shock of compression ratio $R=2.5$ produces a non-thermal synchrotron spectrum of slope ($\alpha+1$) which explains 
the flat part ($\alpha+1=0$ for $R=2.5$) of the observed spectrum. 
This is consistent with the previous finding that at this $R$, the outflow rate is also most significant (Chakrabarti, 1999).
The sharp cut-off in the synchrotron spectrum is due to cut-off in the non-thermal distribution of electrons 
and it is determined by the value of $\gamma$ mentioned above. 
The synchrotron self-Comptonized spectrum due to non-thermal electrons matches with the Chandra data which measures the
core luminosity very accurately. We find the total bolometric luminosity of the nucleus as $9\times10^{42}$ ergs/sec.
Most of the model of M87 are based on the emission from the jet only but we have shown that the spectrum from the nucleus
can be understood by an accretion disk model with a contribution from the jet in the radio range. The acceleration 
of electrons to a relativistic energy and the role of the relativistic electrons in producing the high energy emission 
from the jet may be important to study the spectra from the knots.    
  
\section{Concluding remarks}

In the {\it Letter}, we calculated the spectra from generalized two-component advective accretion disks
located around  stellar mass and super-massive black holes. Specifically, we applied our method for the AGN M87,
perhaps the most massive black hole candidate known. We find that only the sub-Keplerian 
component is enough to describe the black hole spectrum very satisfactorily.
In a sub-Keplerian disk, the flow is almost freely falling and the infall time 
scale for M87 in the CENBOL region ($\sim$ tens of $r_g$) is order of few months. Not surprisingly, the observed
variability of the core of M87 in optical/X-ray wavelength is reported to be of the order of a few months 
(Perlman et al. 2003; Harris, Biretta \& Junor 1997), supporting our view. For a Keplerian disk
the viscous time scale would be a few orders of magnitude higher. Out fit
with a sub-Keplerian flow alone is excellent. What
is more, since the shocks are produced in sub-Keplerian flows, the shock-acceleration provides a natural explanation
for the flat spectrum. The jets which are produced from CENBOL also contribute to the radio emission.
The non-thermal tail due to Comptonization of the synchrotron photons fits the Chandra data and we
believe that it extends at least a few tens of MeV. At higher energies, one may have to consider the hadronic 
interactions which is beyond the scope of the present {\it Letter}.

\acknowledgments  
We are thankful to an anonymous referee for making very constructive comments.
This work is supported by the DST Fast Track Young Scientist Project (SR/FTP/PS-21/2006).

{}

\clearpage
\begin{deluxetable}{cccccc}
\tabletypesize{\scriptsize} \tablecaption{Data for the Nucleus of M87. \label{tbl -1}}
\tablewidth{0pt} \tablehead {\colhead{$\nu$} & \colhead{$\nu F_{\nu}$}
& \colhead{Resolution} & \colhead{Instrument} & \colhead{Observation Year} & \colhead{Reference$^d$}\\ 
Log(Hz) & Log(Jy Hz) & (\arcsec) } 
\startdata
9.166         & $9.732\pm 0.016$   & 1.2    &VLA                       &1985 March 1  & 1\\
9.22          & $9.217^a$          & 0.005  &VLBI                      &1984 April 6  & 2,3\\
9.689         & $10.22\pm 0.01$    & 1.2    &VLA                       &1982 March 2  & 1\\
10.17         & $10.65\pm 0.01$    & 1.2    &VLA                       &1982 March 2  & 1\\
10.34         & $9.681^a$          & 0.00015&VLBI                      & 1986         & 2,4\\
11.0          & $10.94^a$          & 0.0001 & VLBI                     &1989 March 23 & 2,5 \\
13.44         & $11.67\pm 0.02$    & 0.5    & Gemini                   &2001 May 3    & 6\\
14.13         & $11.49\pm 0.1$     & 0.291  &UKIRT                     &1994 June 4   & 7\\
14.378        & $11.767\pm 0.103$  & 0.27   &ESO/MPI                   &1993 May 31   & 7\\
14.78         & $11.79\pm 0.08^b$  & 0.022  &FOC, HST                  & 1991         & 8\\
14.91         & $11.57\pm 0.08^b$  & 0.022  &FOC, HST                  & 1991         & 8\\
14.958        & $11.204\pm 0.004$  & 0.0284 &ACS, HST                  &2003 March 31 & 9\\
15.07         & $11.146\pm 0.003$  & 0.0284 &ACS, HST                  &2003 March 31 & 9\\
15.11         & $11.31\pm 0.08^b$  & 0.022  &FOC, HST                  & 1991         & 8\\
15.11         & $11.48^a$          & 0.05   & FOC, HST                 &1991 April 5  & 2,10 \\
15.28         & $11.11\pm 0.079^b$ & 0.022  &FOC, HST                  & 1991         & 8\\
15.30         & $11.21\pm 0.079^b$ & 0.022  &FOC, HST                  & 1991         & 8\\
15.38         & $11.59\pm 0.079^b$ & 0.022  &FOC, HST                  & 1991         & 8\\
16.68         & $10.53\pm 0.06^c$  & 0.54   &Chandra                   &2000 July 20  & 11\\
17.38         & $10.92\pm 0.04$    & 4.0    &HRI, Einstein Observatory &1979 July 5   & 1\\
18.38         & $10.24\pm 0.08^c$  & 0.54   &Chandra                   &2000 July 20  & 11\\
\enddata
\tablenotetext{a}{Reanalyzed flux densities (ref. 2) from the data given in corresponding cross references 
(ref. 3, 4, 5, 10 respectively) and no mention of error-bars.}
\tablenotetext{b}{The formal flux uncertainty of 20\% has been converted into error-bar.}
\tablenotetext{c}{The flux in the frequency range ($4.8\times10^{16}-2.4\times10^{18}$) Hz has been calculated from the 
flux ($1.07\times10^{-7}$ Jy) at 1 Kev with a slope $\alpha=1.23 \pm 0.11$.}
\tablenotetext{d}{References:- (1) Biretta et a. 1991. (2) Ho 1999. (3) Reid et al. 1989 (4) Spencer \& Junor 1986. 
(5) B\"{a}\"{a}th et al. 1992. (6) Perlman et al. 2001. (7) Stiavelli et al. 1997. (8) Sparks et al. 1996.
(9) Maoz et al. 2005. (10) Maoz et al. 1996. (11) Perlman et al. 2005.}

\end{deluxetable}

\clearpage

\vfil\eject
\begin{figure}
\plotone{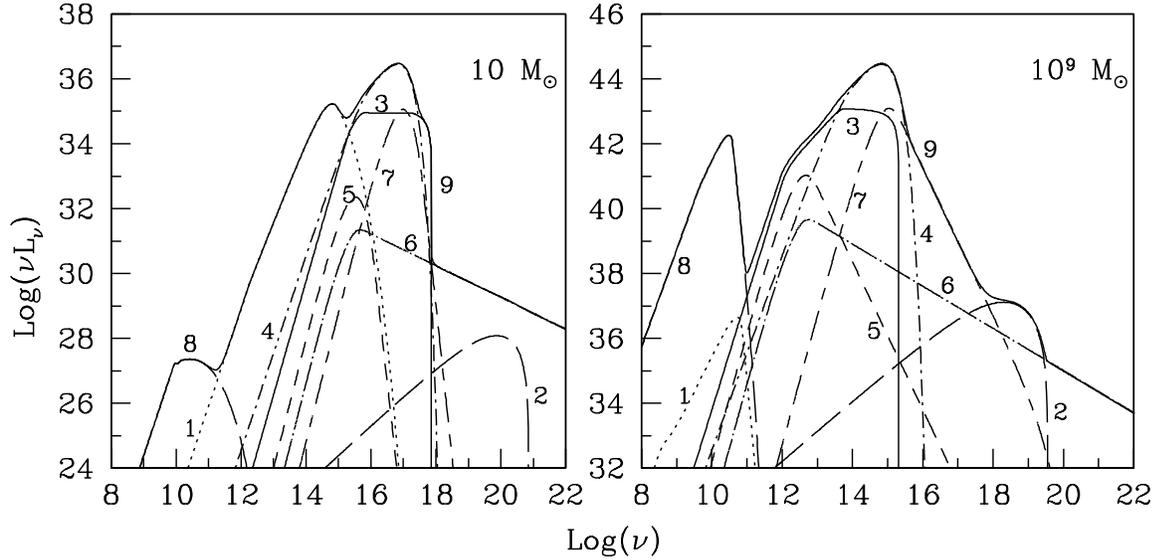}
\caption{The general nature of the two-component advective disk spectrum.
The curves are drawn for different black hole masses (marked) with a same set of flow
parameters. The value of the flow parameters are given in the text. The meaning of the curves are:
1 (dotted) - synchrotron radiation from the pre-shock flow,
2 (long dashed) - bremsstrahlung radiation from the pre-shock flow,
3 (solid) - synchrotron radiation from non-thermal electrons in the CENBOL,
4 (dot-dashed) - black body radiation from the Keplerian disk,
5 (small dashed) \& 6 (dot-long dashed) - Comptonization of the synchrotron soft photons emitted by the
thermal and non-thermal electrons respectively,
7 (short and long dashed) - Comptonization of intercepted black body photons by CENBOL,
8 (long dashed) - synchrotron emission from the jet,
9 (solid) - the total spectrum.}
\end{figure}

\vfil\eject
\begin{figure}
\plotone{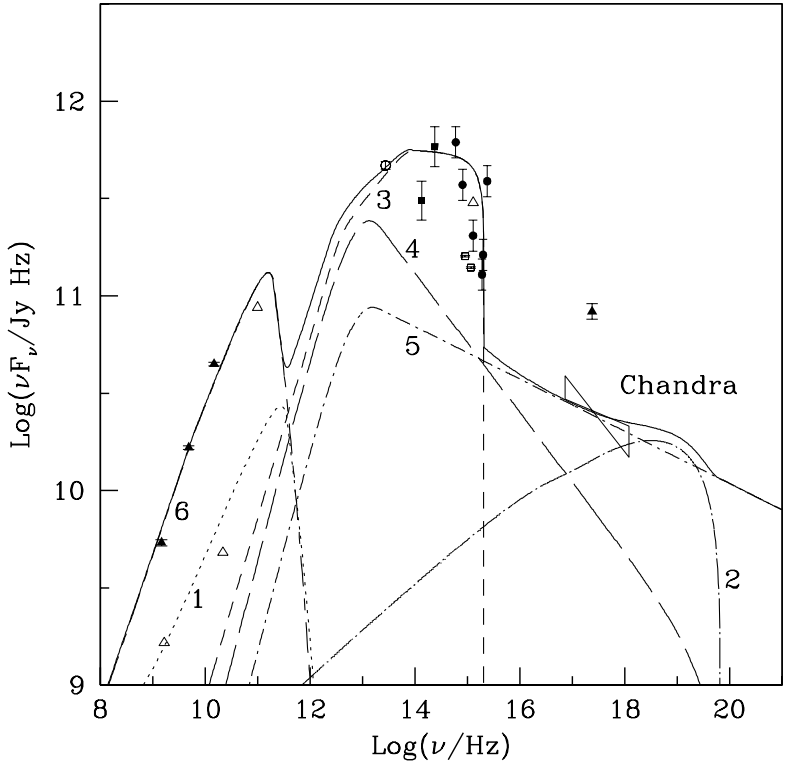}
\caption{Fitting the spectrum of M87 nucleus using the shock solution of sub-Keplerian flow.
The observational data are taken from: Biretta et al. 1991--(filled triangles); Ho 1999 --(open triangles);
Perlman et al. 2001--(open circle); Stiavelli et al. 1997--(filled squares); Sparks et al. 1996--(filled circles);
Maoz et. al 2005--(open squares); Perlman 2005--(Chandra). The dotted line (1) represents
the pre-shock synchrotron contribution while the dot-long-dashed line (2) represents the bremsstrahlung
contribution from the pre-shock flow. The short-dashed line (3) is due to post-shock synchrotron
contribution from non-thermal electrons. The long-dashed (4) and dot-dashed line (5) represents the synchrotron
self-comptonized spectrum due to thermal and non-thermal electrons in the CENBOL.
The long-short-dashed line (6) represents the synchrotron contribution from the jet.}
\end{figure}

\end{document}